\documentclass{CSML}

\def\dOi{11(4:8)2015}
\lmcsheading%
{\dOi}
{1--13}
{}
{}
{Jan.~\phantom06, 2015}
{Dec.~11, 2015}
{}

\ACMCCS{[{\bf Theory of computation}]: Logic---Finite Model Theory; [{\bf Mathematics of
    computing}]: Discrete mathematics}

\subjclass{F.4.1 [Theory of Computation]: Mathematical Logic and Formal
  Languages--Mathematical
Logic; G.2.0 [Mathematics of Computing]: Discrete Mathematics--General}

\pdfoutput=1

\usepackage{hyperref}

\newcommand{\ccc}[1]{{\mbox{\textnormal{\textbf{#1}}}}}  
\newcommand{\cocc}[1]{{\mbox{\textrm{co}\textbf{#1}}}}  

\newcommand{\NP}{\ccc{NP}}

\newcommand{\coNP}{\cocc{NP}}

\newcommand{\FPT}{\ccc{FPT}}

\newcommand{\NOPOLYKERNEL}{{\textnormal\coNP \subseteq \NP/\textup{poly}}}

\newcommand{\PPP}{\mathcal{P}}
\newcommand{\QQQ}{\mathcal{Q}}
\newcommand{\CCC}{\mathcal{C}}
\newcommand{\III}{\mathcal{I}}
\def\ca#1{{\mathcal#1}}

\newcommand{\hy}{\hbox{-}\nobreak\hskip0pt}

\newcommand{\card}[1]{{|#1|}}		
\newcommand{\st}{\ |\ }		     	

\newcommand{\SB}{\{\,}%
\newcommand{\SM}{\;{|}\;}%
\newcommand{\SE}{\,\}}%
\newcommand{\Card}[1]{\card{#1}}

\newcommand{\cov}{\lhd}
\newcommand{\incom}{\parallel}
\newcommand{\width}{\textup{width}}

\newcommand{\cspi}{I}

\newcommand{\tuple}[1]{\langle{#1}\rangle}  

\newcommand{\Fun}{\textup{Pol}}

\newcommand{\defparproblem}[4]{
  \vspace{4pt}
\noindent\fbox{
  \begin{minipage}{0.96\columnwidth}
  \begin{tabular*}{\textwidth}{@{\extracolsep{\fill}}lr} #1  & {\bf{Parameter:}} #3 \\ \end{tabular*}
  {\bf{Input:}} #2  \\
  {\bf{Question:}} #4
  \end{minipage}
  }
  \vspace{4pt}
}

\begin{document}

\title[Faster Existential FO Model Checking on Posets]
      {Faster Existential FO Model Checking on Posets\rsuper*}
\author[J.~Gajarsk\'y]{Jakub Gajarsk\'{y}\rsuper a}
\address{{\lsuper{a,b,c,d}}Faculty of Informatics, Masaryk University, Botanick\'a 68a,
    Brno, 62100, Czech Republic}
\email{\{gajarsky, hlineny, obdrzalek, ordyniak\}@fi.muni.cz}
\thanks{{\lsuper{a,b,c}}Research funded by the Czech Science Foundation under grant
14-03501S}

\author[P.~Hlin\v{e}n\'{y}]{Petr Hlin\v{e}n\'{y}\rsuper b}
\address{\vspace{-18 pt}}

\author[J.~Obdr\v{z}\'{a}lek]{Jan Obdr\v{z}\'{a}lek\rsuper c}
\address{\vspace{-18 pt}}

\author[S.~Ordyniak]{Sebastian Ordyniak\rsuper d}
\address{\vspace{-18 pt}}
\thanks{{\lsuper d}Research funded by 
      Employment of Newly Graduated Doctors of Science for
      Scientific Excellence (CZ.1.07/2.3.00/30.0009). %
      \emph{Current address:} Institute of Information Systems, Vienna University of
  Technology, Favoritenstrasse 9-11, A-1040 Vienna, Austria}

\keywords{first-order logic; partially ordered sets; model checking;
  parameterized complexity}
\titlecomment{{\lsuper*}An extended abstract of an earlier version of this paper
  has appeared at ISAAC'14.}

\begin{abstract}
  We prove that the model checking problem for the existential fragment of
  first-order (FO) logic on partially ordered sets is fixed-parameter
  tractable (FPT) with respect to the formula and the width of a poset (the
  maximum size of an antichain).  While there is a long line of research
  into FO model checking on graphs, the study of this problem on posets has
  been initiated just recently by Bova, Ganian and Szeider (CSL-LICS 2014), who
  proved that the existential fragment of FO has an FPT algorithm for a
  poset of fixed width.  We improve upon their result in two ways: (1) the runtime of
  our algorithm is $O(f(|\phi|,w)\cdot n^2)$ on $n$-element posets of width
  $w$, compared to $O(g(|\phi|)\cdot n^{h(w)})$ of Bova et al., and (2) our
  proofs are simpler and easier to follow.  We complement this result by
  showing that, under a certain complexity-theoretical assumption, the
  existential FO model checking problem does not have a polynomial kernel.
\end{abstract}

\maketitle

\section{Introduction}

The {\em model checking} problem, asking whether a logical formula holds true on a
given input structure, is a fundamental problem of theoretical computer science with
applications in many different areas, e.g. algorithm design or formal
verification. 
One way to see why providing efficient algorithms for model
checking is important is to note that such algorithms automatically establish
efficient solvability of whole classes of problems.
For first-order (FO) logic, the model checking problem is known to be
PSPACE-complete when the formula is part of the input, and polynomial time
solvable when the formula is fixed in advance.

However, this does not tell the whole story. 
In the latter scenario we would like to identify the
instances where we could do significantly better---in regard to
running times---and quantify these gains. Stated in the parlance
of parameterized complexity theory, we wish
to identify classes of input structures on which we can evaluate every FO
formula $\phi$ in polynomial time $f(|\phi|)\cdot n^c$, where $c$ is a constant
independent of the formula. If it is true, we
say that FO model checking problem is \emph{fixed-parameter tractable} (FPT)
on this class of structures. 

Over the past decade this line of research has been very
active  and led to several important results on (mainly) undirected graphs, 
which culminated in the recent result of Grohe, Kreutzer and Siebertz~\cite{gks14}, 
stating that FO model checking is fixed-parameter tractable on all nowhere dense classes of graphs. 

In contrast,
almost nothing is known about the complexity of FO model checking
on other finite algebraic structures. Very recently, Bova, Ganian and
Szeider~\cite{BGS14} initiated the study of the model checking problem 
for FO and partially ordered sets. 
Despite similarities between posets and graphs (e.g., in Hasse diagrams),
the existing FO model checking results from graphs do not seem to transfer
well to posets, perhaps due to lack of usable notions of 
``locality'' and ``sparsity'' there.
This feeling is supported by several negative results in \cite{BGS14}, too.

The main result of Bova et al.~\cite{BGS14} is that the model checking problem for the existential
fragment of~FO  (\textsc{Poset $\exists$-FO-Model Checking})
can be solved in time $f(|\phi|)\cdot n^{g(w)}$, where $n$ is the size
of a poset and $w$ its {\em width}, i.e. the size of its largest antichain. 
In the language of parameterized complexity, this means
that the problem is FPT in the size of the
formula, but only XP with respect to the width of the poset. 
Note that this is not an easy result since, for instance,
posets of fixed width can have unbounded clique-width~\cite{BGS14}.

The proof in~\cite{BGS14} goes by first showing that the
model checking problem for the existential fragment of FO is 
equivalent to the embedding problem for posets (which can be thought
as analogous to the induced subgraph problem),
and then reducing the embedding problem to a suitable family of instances of
the homomorphism problem of certain semilattice structures. 

While postponing further formal definitions till Section~\ref{sec:prelim},
we now state our main result which improves upon the aforementioned result of Bova et al.:
\begin{thm}\label{thm:existfo-fpt}
  \textsc{Poset $\exists$-FO-Model Checking} is fixed-parameter tractable
  in the formula size and the width of an input poset;
  precisely, solvable in time $h(|\phi|,w)\cdot O(n^2)$ where $n$ is the size
  of a poset and $w$ its width.
\end{thm}

Our improvement is two-fold;
(1) we show that
the existential FO model checking problem is fixed-parameter tractable in
\emph{both} the size of the formula and the width of the poset, and (2)
we give two simpler proofs of this result, one of them completely {\em self-contained}. 
Regarding improvement (2), we use the same reduction of existential FO model checking
to the embedding problem from \cite{BGS14}, but our subsequent
solution to embedding is faster and at the same time
much more straightforward and easier to follow.

As stated above, we give two different FPT algorithms solving the poset embedding problem (and thus also the
existential FO model checking problem). The first algorithm
(Section~\ref{sec:embeddingfpt}) is a natural,
and easy to understand, polynomial-time reduction to a CSP (Constraint
Satisfaction Problem) instance closed
under min polymorphisms, giving us an $O(n^4)$ dependence of the running time
on the size of the poset. 
The second algorithm (Section~\ref{sec:embclique}) has even better, quadratic,
time complexity and works by reducing the embedding problem to a restricted
variant of the multicoloured clique problem, which is then efficiently
solved.

To complement the previous fixed-parameter tractability results, we also
investigate possible kernelization of the embedding problem for posets
(Section~\ref{sec:lowerkernel}). We
show that the embedding problem
does not have a polynomial kernel, unless $\NOPOLYKERNEL$,
which is thought to be unlikely. This means the embedding problem 
(and therefore also the existential and full FO model checking problems)
cannot be efficiently reduced to an equivalent instance of size polynomial
in the parameter.

\section{Preliminaries}
\label{sec:prelim}

\subsection{Posets and Embedding}

A \emph{poset} $\PPP$ is a pair $(P,\leq^P)$ where $P$ is a set and $\leq^P$ is
a reflexive, antisymmetric, and transitive binary relation over $P$.
The \emph{size} of a poset $\PPP=(P,\leq^P)$ is $\Vert\ca P\Vert:=|P|$. 
We say that $p$ \emph{covers} $p'$ for $p,p' \in P$, denoted by $p'
\cov^P p$, if $p' \leq^P\! p$, $p
\neq p'$, and for every $p''$ with $p' \leq^P\! p''\leq^P\! p$ it
holds that $p''\in\{p,p'\}$. We say that $p$ and $p'$ are \emph{incomparable}
(in $\PPP$), denoted $p \incom^P p'$ if neither $p \leq^P\! p'$ nor $p'
\leq^P\! p$. A \emph{chain} $C$ of
$\PPP$ is a subset of $P$ such that $x \leq^P y$ or $y \leq^P x$
for every $x,y \in C$. An \emph{anti-chain} $A$ of $\PPP$ is a subset
of $P$ such that for all $x,y \in P$ it is true that  $x \incom^P y$.
A \emph{chain partition} of $\PPP$ is a tuple $(C_1,\dotsc,C_k)$ such
that $\{C_1,\dotsc,C_k\}$ is a partition of $P$ and for every $i$ with
$1 \leq i \leq k$ the poset induced by $C_i$ is a chain of $\PPP$.
The \emph{width} of a poset $\PPP$, denoted by $\width(\PPP)$ is the 
maximum cardinality of any anti-chain of $\PPP$.
\begin{prop}[{\cite[Theorem 1.]{frs03}}]\label{pro:comp-chain-part}
  Let $\PPP$ be a poset. Then in time $O(\width(\PPP)\cdot\Vert\PPP\Vert^2)$, it is
  possible to compute both $\width(\PPP)=w$ and a
  corresponding chain partition $(C_1,\dotsc,C_{w})$ of $\PPP$.
\end{prop}

Let $\QQQ=(Q, \leq^Q)$ and $\PPP=(P, \leq^P)$ be two posets.
An \emph{embedding} from $\QQQ$ to $\PPP$ is an injective 
function $e : Q \rightarrow P$ such that, $q \leq^Q\! q'$ if and only if
$e(q) \leq^P\! e(q')$ for every $q,q' \in Q$. The \emph{embedding problem}
for posets is thus defined as: 

\defparproblem{{\sc Embedding}}{Two posets $\QQQ=(Q,\leq^Q)$ and
$\PPP=(P,\leq^P)$.}{$\width(\PPP)$, $\Vert\QQQ\Vert$}
{Is there an embedding from $\QQQ$ into $\PPP$?}

\subsection{Constraint Satisfaction Problems}

A \emph{constraint satisfaction problem} (CSP) $\cspi$
is a triple $\tuple{V,D,C}$, where $V$ is a finite set of variables
over a finite set (domain) $D$, and $C$ is a set of constraints. A
\emph{constraint} $c \in C$ consists of a \emph{scope},
denoted by $V(c)$, which is an ordered subset of
$V$, and a relation, denoted by $R(c)$, which is a
$|V(c)|$-ary relation on $D$. 
For a CSP $\cspi=\tuple{V,D,C}$ we sometimes denote by $V(\cspi)$,
$D(\cspi)$, and $C(\cspi)$, its set of variables $V$, its domain $D$,
and its set of constraints $C$, respectively. 
A \emph{solution} to a CSP instance $\cspi$ is a mapping $\tau : V
\rightarrow D$ such that $\tuple{\tau[v_1],\dotsc,\tau[v_{|V(c)|}]}
\in R(c)$ for every $c \in C$ with $V(c)=\tuple{v_1,\dotsc,v_{|V(c)|}}$.

Given a $k$-ary relation $R$ over some domain $D$ and a function
$\phi:D^n \rightarrow D$, we say that $R$ is {\em closed under
  $\phi$}, if for all collections of $n$ tuples $t_1,\dotsc,t_n$ from
$R$, the tuple $\tuple{\phi(t_1[1],\dotsc,t_n[1]),
  \dotsc,\phi(t_1[k],\dotsc,t_n[k])}$ belongs to $R$. The function
$\phi$ is also said to be a {\em polymorphism of $R$}. 
We denote by $\Fun(R)$ the set of all polymorphisms $\phi$ such that
$R$ is closed under $\phi$.

Let $\cspi=\tuple{V,D,C}$ be a CSP instance and $c \in C$.  We write
$\Fun(c)$ for the set $\Fun(R(c))$ and we write $\Fun(\cspi)$ for the
set $\bigcap_{c \in C}\Fun(c)$.  We say that $\cspi$ is closed under a
polymorphism $\phi$ if $\phi \in \Fun(\cspi)$.

We will need the following type of polymorphism. A polymorphism $\phi
: D^2 \rightarrow D$ is a
\emph{min} polymorphism if there is an ordering of the
elements of $D$ such that for every $d,d' \in D$, it holds that
$\phi(d,d')=\phi(d',d)=\min\{d,d'\}$.

\begin{prop}[{\cite[Corollary 4.3]{jcg97}}]\label{pro:solve-csp}
  Any CSP instance $\cspi$ that is closed under a min polymorphism 
  (that is provided with the input) can be
  solved in time $O((ct)^2)$, where $c=|C(\cspi)|$ and $t$ is the maximum cardinality
  of any constraint relation of $\cspi$.
\end{prop}

\subsection{Parameterized Complexity}

Here we introduce the relevant concepts of parameterized complexity theory.
For more details, we refer to text books on the topic~\cite{df99,fg06,nie06}.
An instance of a parameterized problem is a pair $\tuple{x,k}$ where $x$ is
the input and $k$ a parameter.  A~parameterized 
problem is \emph{fixed-parameter tractable} if every instance
$\tuple{x,k}$ can be solved in time $f(k)\cdot\Card{x}^c$, where $f$ is a
computable function, and $c$ is a constant.
$\FPT$ denotes the class of all fixed-parameter tractable problems.  

A \emph{kernelization} \cite{agksy11} 
for a parameterized problem~$\ca A$ is a polynomial time
algorithm that takes an instance
$\tuple{x,k}$ of $\ca A$ and maps it to an equivalent 
instance $\tuple{x',k'}$ of 
$\ca A$ such that
both $|x'|$ and $k'$ are bounded by some function $f$~of~$k$.  
The output $\tuple{x',k'}$ is called a \emph{kernel}. 
We say that $\ca A$
has a \emph{polynomial kernel} if $f$ is a polynomial.  
Every fixed-parameter tractable problem admits a kernel, 
but not necessarily a polynomial kernel~\cite{Caietal93}.

A \emph{polynomial parameter reduction} from a parameterized problem $\ca A$
to a parameterized problem $\ca B$ 
is a polynomial time algorithm, which, given an instance $\tuple{x,k}$ of
$\ca A$ produces an instance $\tuple{x',k'}$ of $\ca B$ such that
$\tuple{x,k}$ is a \textsc{Yes}-instance of
$\ca A$ if and only if $\tuple{x',k'}$ is a \textsc{Yes}-instance of
$\ca B$ and $k'$ is bounded by some polynomial of~$k$.  
The following results show how polynomial parameter
reductions can be employed to prove the non-existence of polynomial
kernels.
\begin{prop}[{\cite[Theorem~8]{bod09}}]\label{pro:poly-par-reduction-bi}
  Let $\ca A$ and $\ca B$ be two parameterized problems such that there is a
  polynomial parameter reduction from $\ca A$ to $\ca B$. If $\ca B$
  has a polynomial kernel, then so has~$\ca A$.
\end{prop}

An \emph{OR-composition algorithm} for a parameterized problem $\ca A$ maps
any $t$ instances $\tuple{x_1,k},\dotsc,\tuple{x_t,k}$ of $\ca A$ to
one instance $\tuple{x',k'}$ of $\ca A$ such that the algorithm runs in
time polynomial in $\sum_{1 \leq i \leq t}|x_i|+k$, the parameter
$k'$ is bounded by a polynomial in the parameter $k$, and
$\tuple{x',k'}$ is a \textsc{Yes}-instance if and only if 
there exists $1 \leq i \leq t$ such that 
$\tuple{x_i,k}$ is a \textsc{Yes}-instance.
\begin{prop}[{\cite[Lemmas~1 and~2]{bdfh09}}]\label{pro:or-comp-no-poly-kernel}
  If a parameterized problem $\ca A$ has an OR\hy composition algorithm
  and its unparameterized version is \NP-complete,
  then $\ca A$ has no polynomial kernel, unless $\NOPOLYKERNEL$.
\end{prop}

\subsection{Existential First-order Logic}

In this paper we deal with relational first-order (FO)
logic. Formulas of this logic are built from (a countable set of) variables,
relational symbols, logical connectives ($\land, \lor, \neg$) and
quantifiers ($\exists, \forall$). A sentence is a formula with no free
variables. We restrict ourselves to formulas that are in \emph{prefix normal
  form}. (A first-order formula is in prefix normal form if all quantifiers
occur in front of the formula and all negations occur in front of the
atoms.) Furthermore an \emph{existential} first-order formula is a first-order formula
in prefix normal form that uses only existential quantifiers. 

The problem we are interested in is the so-called \emph{model checking problem} for the
existential FO formulas (and posets), which is formally defined as follows:

\defparproblem{{\sc Poset $\exists$-FO-Model Checking}}{An existential first-order sentence $\phi$ and a poset $\PPP=(P,\leq^P)$.}
{$\width(\ca P)$, $|\phi|$}{Is it true $\PPP\models\phi$, i.e., is $\PPP$ a model of $\phi$?}

We remark here that all first-order formulas in this paper are
evaluated over posets. In particular, the vocabulary of these formulas
consists of only one binary relation $\leq$ and atoms of these
formulas can be either equalities between variables ($x=y$) or
applications of the predicate $\leq$ ($x\leq\! y$). (Which is, of course,
interpreted by $\leq^P$ for a concrete poset $\ca P$.)
For a more detailed treatment of the employed setting, we refer the reader
to~\cite{BGS14}.

As shown in~\cite{BGS14}, the existential FO model checking problem is
closely related to the aforementioned \emph{embedding problem} for posets: 

\begin{prop}[\cite{BGS14}]
\label{pro:equiv-FO-Embedding}
  \textsc{Poset $\exists$-FO-Model Checking} is fixed-parameter tractable if and only if
  so is \textsc{Embedding}. Moreover, there is a polynomial parameter reduction from
  \textsc{Embedding} to \textsc{Poset $\exists$-FO-Model Checking}.
\end{prop}
\proof
  The first statement of the proposition follows immediately from~\cite[Proposition 1]{BGS14}.
  The second statement of the proposition follows from the proof of~\cite[Proposition 1]{BGS14}
  by observing that the obvious reduction from \textsc{Embedding} to \textsc{Poset $\exists$-FO-Model Checking}
  is polynomial parameter preserving.
\qed

\begin{rem}\label{rem:equiv-FO-Embedding}
Even though \cite{BGS14} does not state the precise runtime and 
``instance blow-up''
for Proposition~\ref{pro:equiv-FO-Embedding},
these can be alternatively bounded from above as follows.
For an instance $(\PPP,\phi)$ where $\phi\equiv\exists x_1\dots\exists x_q.\,\psi(x_1,\dots,x_q)$,
we exhaustively enumerate all posets $\ca Q$ on $Q=\{x_1,\dots,x_q\}$
(modulo equality $=$ on $Q$)
such that $\ca Q\models\psi$, and produce a separate instance of
\textsc{Embedding} from this particular $\ca Q$ into the same $\ca P$.
Then $\ca P\models\phi$ if and only if at least one of the constructed
\textsc{Embedding} instances is {\sc Yes}. 
The number of produced instances (of $\ca Q$) 
is trivially less than the number of all posets on $q$ elements factorized by
equality, $<4^{q^2}=2^{O(|\phi|^2)}$,
and time spent per each one of them in the construction is~$O(|\phi|^2)$.
\end{rem}

\section{Fixed-parameter Tractability Proof}
\label{sec:embeddingfpt}

In this section we prove the first half of the main result of our
paper (The\-orem~\ref{thm:existfo-fpt})
that the existential FO model checking problem for posets is in FPT.
By Proposition~\ref{pro:equiv-FO-Embedding}, it is enough to
consider the embedding problem for that:

\begin{thm}\label{the:embedding-fpt}
  Let $\QQQ=(Q,\leq^Q)$ and $\PPP=(P,\leq^P)$ be two posets. Then the
  embedding problem from $\QQQ$ into $\PPP$ is fixed-parameter tractable,
  more precisely, it can be
  solved in time $O\big(\width(\PPP)^{|Q|}\cdot|Q|^4\cdot|P|^4\big)$.
\end{thm}

The remainder of this section is devoted to a proof of the above
theorem. Let $w:=\width(\PPP)$ for the rest of this section. The algorithm
starts by computing a chain partition $\CCC=(C_1,\dotsc,C_w)$ of $\PPP$.
This can be done in time $O(\width(\PPP)\cdot|P|^2)$
by Proposition~\ref{pro:comp-chain-part}.

To make the proof clearer, we will, for an embedding, keep track into which
chain each element of $\QQQ$ is mapped. We say that an embedding $e$ from
$\QQQ$ into~$\PPP$ is \emph{compatible} with a function $f$ from $Q$ to $\{1, \dotsc,
w\}$ if $e(q) \in C_{f(q)}$ for every $q \in Q$.  Observe that every embedding
$e$ is trivially compatible with the unique function $f$, where $f(q)=i$ if
and only if $e(q) \in C_i$. Also note that there are at most
$(\width(\PPP)^{|Q|})$ such functions $f$.

Our algorithm now will do the following: We generate all possible functions
$f$ (as defined in the previous paragraph) and for each such $f$ we test
whether there is an embedding compatible with $f$. The following lemma,
stating that we can perform such a test efficiently, forms the core of our proof.

\begin{lem}\label{lem:compatible-embedding}
  Let $f$ be a function from $Q$ to $\{1,\dotsc,w\}$ where $w=\width(\PPP)$. Then
  one can decide in time $O\big(|Q|^4\cdot|P|^4\big)$ whether there is an embedding $e$
  from $\QQQ$ to $\PPP$ that is compatible with $f$.
\end{lem}
\proof
We will prove the lemma by reducing the problem (of finding a
compatible embedding) in polynomial time to a CSP instance that
is closed under a certain min polymorphism and hence can be solved in
polynomial time. We start by defining the CSP instance $\cspi$ for
given $\QQQ$, $\PPP$, $f$, and $\CCC$ as above. 

$\cspi$ has one variable $x_q$ for every $q \in Q$ whose domain are
the elements of $C_{f(q)}$.
Furthermore, for every pair $q,q'$ of distinct elements of
$Q$, $\cspi$ contains one constraint $c_{q,q'}$ whose scope is $(x_q,x_{q'})$
and whose relation $R(c_{q,q'})$ contains all tuples $(p,p')$ such 
that $p \in C_{f(q)}$, $p' \in C_{f(q')}$, and simultaneously
\begin{enumerate}[label=\arabic*.]
\item $p \leq^P\! p'$ iff $q \leq^Q\! q'$,
\item $p' \leq^P\! p$, iff $q' \leq^Q\! q$.
\end{enumerate}
This completes the construction of $\cspi$. 
Observe that a solution $\tau : V(\cspi) \rightarrow D(\cspi)$ of
$\cspi$ gives rise to an embedding $e : Q \rightarrow P$ from $\QQQ$
to $\PPP$ that is compatible with $f$ by setting
$e(q)=\tau(x_q)$. Additionally, every embedding $e : Q \rightarrow P$
from $\QQQ$ to $\PPP$ that is compatible with $f$ gives rise to a
solution $\tau : V(\cspi) \rightarrow D(\cspi)$ of $\cspi$ by setting
$\tau(x_q)=e(q)$. Hence, $\cspi$ has a solution if and only if 
there is an embedding from $\QQQ$ to $\PPP$ that is compatible with
$f$ and such an embedding can be easily obtained from a solution of
$\cspi$.

Concerning the runtime, $\cspi$ can be constructed in
time $O((|Q|\cdot|P|)^2)$.
Since there are less than $|Q|^2$ constraints and every constraint relation
contains $O(|P|^2)$ pairs, Proposition~\ref{pro:solve-csp} provides a
solution to $I$ in time $O((|Q|^2\cdot|P|^2)^2)$.
To finish it is enough to verify
that $\cspi$ is closed under a certain min polymorphism---%
Lemma~\ref{lem:polymorphism} below.
\qed

\begin{lem}
\label{lem:polymorphism}
  For every $\QQQ$, $\PPP$, $f$, and $\CCC$ defined as above, the CSP
  instance $\cspi$ is closed under any min polymorphism that is
  compatible with the partial order~$\leq^P$.
\end{lem}
\proof  
  In the following, let $c_{q,q'}$ be a constraint of $\cspi$ for two
  distinct elements $q,q'
  \in Q$ and let $(p_1,p_2) \in R(c_{q,q'})$ and $(p_1',p_2') \in
  R(c_{q,q'})$. We need to show 
  $(\min_{\leq^P}\{p_1,p_1'\},\min_{\leq^P}\{p_2,p_2'\}) \in
  R(c_{q,q'})$. Observe here and in the following that
  $\min_{\leq^P}\{p_1,p_1'\}$ and $\min_{\leq^P}\{p_2,p_2'\}$ are
  well-defined because $p_1$ and $p_1'$ and $p_2$ and $p_2'$ both lie in
  $C_{f(q)}$ and $C_{f(q')}$, respectively.
  We distinguish three cases (depending on the relationship of $q$ and
  $q'$ with respect to $\leq^Q$):

  \begin{enumerate}
  \item If $q <^Q q'$, then by the definition of $\cspi$, the
    relation $R(c_{q,q'})$ contains all tuples $(p,p')$ such that $p \in
    C_{f(q)}$, $p' \in C_{f(q')}$, and $p <^P p'$. It follows that
    $p_1 <^P p_2$ and $p_1' <^P p_2'$. Hence,
    $\min_{\leq^P}\{p_1,p_1'\} <^P \min_{\leq^P}\{p_2,p_2'\}$ (by
    transitivity of $\leq^P$) and
    consequently 
    $\big(\min_{\leq^P}\{p_1,p_1'\},\min_{\leq^P}\{p_2,p_2'\}\big) \in
    R(c_{q,q'})$, as required. 
  \item The case that $q' <^Q q$ is symmetric to the previous case.
  \item If $q \incom^Q q'$, then by the definition of $\cspi$, the
    relation $R(c_{q,q'})$ contains all tuples $(p,p')$ such that $p \in
    C_{f(q)}$, $p' \in C_{f(q')}$, and $p \incom^P p'$. It follows that
    $p_1 \incom^P p_2$ and $p_1' \incom^P p_2'$. Clearly, if
    $\big(\min_{\leq^P}\{p_1,p_1'\},\min_{\leq^P}\{p_2,p_2'\}\big) \in
    \{(p_1,p_2),(p_1',p_2')\}$, then there is nothing to show. Hence,
    assume that this is not the case and assume w.l.o.g. that $p_1
    \leq^P p_1'$. Then,
    $\big(\min_{\leq^P}\{p_1,p_1'\},\min_{\leq^P}\{p_2,p_2'\}\big)=(p_1,p_2')$.
    If $p_1 \leq^P p_2'$, then because $p_2' \leq^P p_2$ also $p_1
    \leq^P p_2$, a contradiction to our assumption that $p_1 \incom^P
    p_2$. Similarly, if $p_2' \leq^P p_1$, 
    then because $p_1 \leq^P p_1'$ also $p_2'
    \leq^P p_1'$, a contradiction to our assumption that $p_1' \incom^P
    p_2'$. Hence,
    $\min_{\leq^P}\{p_1,p_1'\} \incom^P \min_{\leq^P}\{p_2,p_2'\}$ and
    consequently 
    $\big(\min_{\leq^P}\{p_1,p_1'\},\min_{\leq^P}\{p_2,p_2'\}\big) \in
    R(c_{q,q'})$, as required.
 \qed
  \end{enumerate}

\proof[Proof of Theorem~\ref{the:embedding-fpt}]
We can generate the chain partition in time
$O(\width(\PPP)\cdot|P|^2)$. Then, for each of the $(\width(\PPP)^{|Q|})$
functions $f$ we test the existence of an embedding compatible with $f$,
which can be done in time $O\big(|Q|^4\cdot|P|^4\big)$ by
Lemma~\ref{lem:compatible-embedding}. This proves our theorem.
\qed

\section{Embedding and Multicoloured Clique}
\label{sec:embclique}
 
In the previous section we have proved that the embedding problem for posets
$\QQQ$ and $\PPP$ is fixed-parameter tractable w.r.t. both $\width(\PPP)$
and $\Vert\QQQ\Vert$, with the running time of
$O\big(\width(\PPP)^{\Vert\QQQ\Vert}\cdot\Vert\QQQ\Vert^4\cdot\Vert\PPP\Vert^4\big)$. 
In this section we
improve upon this result by giving an alternative self-contained algorithm for
\textsc{Embedding} with running time
$O\big(\width(\PPP)^{\Vert\QQQ\Vert}\cdot\Vert\QQQ\Vert^3\cdot\Vert\PPP\Vert^2\big)$. 
In combination with Proposition~\ref{pro:equiv-FO-Embedding}
(and Remark~\ref{rem:equiv-FO-Embedding})
we thus finish the proof of main Theorem~\ref{thm:existfo-fpt}.

This new algorithm achieves better efficiency by
exploiting some special properties of the problem that are not fully
utilized in the previous reduction to CSP. We pay for this improvement by
having to work a little bit harder.
The core idea is to show that the problem of finding a
compatible embedding is reducible (in polynomial time) to a certain
restricted variant of \textsc{Multicoloured Clique}.  

\defparproblem{{\sc Multicoloured Clique}}
{A graph $G$ with a proper $k$-colouring of its vertices.}
{$k$}{Is there a clique (set of pairwise adjacent vertices) of size $k$ in $G$?}

The \textsc{Multicoloured Clique} problem takes as an input a graph $G$ together
with a proper $k$-colouring of the vertices of~$G$.
The question is whether there is a $k$-clique in $G$. (Note that the
vertices of a clique in a properly coloured graph necessarily get distinct colours.)

Consider posets $\ca Q=(Q,\leq^Q)$, $\ca P=(P, \leq^P)$ and
a chain partition $(C_1,\dots,C_w)$ of $\ca P=(P, \leq^P)$
where $w=\width(\ca P)$.
Let $f:Q\to\{1,\dots,w\}$ be an arbitrary function and, for
simplicity, assume $Q=\{1,\dots,k\}$.
We construct a $k$-coloured graph $G=G(\ca P,\ca Q,f)$ as follows.
The vertex set of $G$ is a disjoint union $V(G)=V_1\,\dot\cup\dots\dot\cup\,V_k$
of $k$ colour classes where $V_i$, $i\in Q$, is a copy of $C_{f(i)}$.
Let $i,j\in Q$ and let $p\in V_i, q\in V_j$ be the corresponding copies 
of arbitrary $p'\in C_{f(i)}, q'\in C_{f(j)}$.
Then we put $pq\in E(G)$ if and only if $i\not=j$ and the following hold;
\begin{enumerate}[label=\arabic*.]
\item $p'\leq^P\!q'$ iff $i\leq^Q\!j$, and
\item $p'\geq^P\!q'$ iff $i\geq^Q\!j$.
\end{enumerate}

\begin{prop}
\label{prop:mcclique}
For any two posets $\ca Q=(Q,\leq^Q)$, $\ca P=(P, \leq^P)$,
any chain partition $(C_1,\dots,C_w)$ of $\ca P$,
and arbitrary $f:Q\to\{1,\dots,w\}$
the graph $G(\ca P,\ca Q,f)$ is a \textsc{Yes}-instance of $|Q|$-coloured
{\sc Multicoloured Clique} problem 
if and only if $\ca Q$ has an $f$-compatible embedding into~$\ca P$.
\end{prop}
\proof
Consider a \textsc{Yes}-instance of $G:=G(\ca P,\ca Q,f)$, 
which means there is a clique $K\subseteq V(G)$ of size~$k=|Q|$ 
(and thus intersecting each one of $V_1,\dots,V_k$ of $G$ exactly once).
For $i\in Q$, let the embedding map $i$ to $e(i):=p'\in C_{f(i)}$ such that
$V_i\cap K=\{p\}$ and $p$ is the corresponding copy of~$p'$ in the
construction of~$G$.
Then immediately; $i \leq^Q\! j$ if and only if
$e(i) \leq^P\! e(j)$ for every $i,j \in Q$.

Conversely, consider an $f$-compatible embedding $e:Q\to P$.
We define $K:=\{p: i\in Q \mbox{ and $p\in V_i$ is the copy of } e(i)\}$.
Then $K$ is a clique of size $|Q|$ by the definition of~$G$.
\qed

For reference,
we associate each colour class $V_i$, $i\in Q$, of $G=G(\ca P,\ca Q,f)$ with a 
linear order $\leq^G$ naturally inherited from the corresponding chain of $\ca P$
(we are not going to compare between different classes).

\begin{lem}
\label{lem:minmaxcliq}
Let $G:=G(\ca P,\ca Q,f)$ be as in Proposition~\ref{prop:mcclique}
and $V_i$, $i\in Q$, be the colour classes of~$G$.
Let $i,j\in Q$ be any two elements such that $i\not=j$. Then the following two
statements are true:
\begin{enumerate}[label=\roman*)]
\item
For any $p\in V_i$, $q_1,q_2,q_3\in V_j$ such that 
$q_1\leq^G\!q_2\leq^G\!q_3$ it holds;
if $pq_1,\,pq_3\in E(G)$ then also $pq_2\in E(G)$.
\item
For any $p_1,p_2\in V_i$, $q_1,q_2\in V_j$ such that $p_1\leq^G\!p_2$,
$q_1\leq^G\!q_2$ it holds;
if $p_1q_2,\,p_2q_1\in E(G)$ then also $p_1q_1,\,p_2q_2\in E(G)$.
\end{enumerate}
\end{lem}
\proof
This follows similarly to the arguments from Lemma~\ref{lem:polymorphism}.

a)
Let $p'\in C_{f(i)}$, $q_1',q_2',q_3'\in C_{f(j)}$ be the corresponding
points of $\ca P$, and assume $pq_2\not\in E(G)$.
If $i\leq^Q\!j$, then $p'\not\leq^P\!q_2'$ by~$G$ but
$p'\leq^P\!q_1'\leq^P\!q_2'$ by transitivity in~$\ca P$.
The case $i\geq^Q\!j$ is analogous.
If $i\incom^Q\!j$,
then $p'\incom^P\!q_1'$, $p'\incom^P\!q_3'$ by the definition of $E(G)$,
but $p'\leq^P\!q_2'$ or $p'\geq^P\!q_2'$.
Each of the latter possibilities contradicts transitivity in~$\ca P$.

b)
Let $p_1',p_2'\in C_{f(i)}$, $q_1',q_2'\in C_{f(j)}$ be the corresponding
points of $\ca P$, and assume $p_1q_1\not\in E(G)$.
If $i\leq^Q\!j$, then $p_1'\not\leq^P\!q_1'$ but
$p_1'\leq^P\!p_2'\leq^P\!q_1'$ by the edge $p_2q_1\in E(G)$ and 
transitivity in~$\ca P$, a contradiction.
The case $i\geq^Q\!j$ is analogous.
If $i\incom^Q\!j$
then, up to symmetry, $p_1'\leq^P\!q_1'$ and so $p_1'\leq^P\!q_2'$ by
transitivity in $\ca P$, contradicting assumed
$i\incom^Q\!j$ $\iff$ $p_1'\incom^P\!q_2'$.
\qed

We call a \textsc{Multicoloured Clique} instance $G$ {\em interval-monotone}
if the colour classes of $G$ can be given linear order(s) $\leq^G$ such that
both conditions a),b) as in Lemma~\ref{lem:minmaxcliq} are satisfied.

\begin{cor}
\label{cor:minmaxcliq}
Let $G$ be an interval-monotone (wrt. $\leq^G$)
multicoloured clique instance with colour classes $V_1,\dots,V_k$.
Let $I\subseteq\{1,\dots,k\}$.
If $K_1,\dots,K_\ell\subseteq \bigcup_{i\in I}V_i$ are cliques of size
$|I|$, then also the set
$$ K=\big\{\min\nolimits_{\leq^G}\big((K_1\cup\dots\cup K_\ell)\cap V_i\big)
	\st i\in I \big\}, $$
called the {\em minimum of $K_1,\dots,K_\ell$} wrt. $\leq^G$ and $I$, is a clique in~$G$.
The same holds for analogous {\em maximum of $K_1,\dots,K_\ell$}
wrt. $\leq^G$ and $I$.
\end{cor}
\proof
Let $K=\{v_i\st i\in I\}$ and $g$ be a function such that
$v_i\in K_{g(i)}\cap V_i$ for all $i\in I$.
If $i\not=j\in I$, then both $v_iw_{i,j},\,v_jw_{j,i}\in E(G)$
where $\{w_{i,j}\}=K_{g(i)}\cap V_j$, by the assumptions.
Clearly, the assumptions of Lemma~\ref{lem:minmaxcliq}\,b) are satisfied
for $v_iw_{i,j},\,v_jw_{j,i}$, and hence $v_iv_j\in E(G)$.
\qed

For smooth explanation of our algorithm, we introduce the following shorthand
notation.
Let $[i,j]=\{i,i+1,\dots,j\}$.
Let $V(G)=V_1\,\dot\cup\dots\dot\cup\,V_k$.
Then $N_i(v)$ denotes the set of neighbours of $v\in V(G)$ in
$V_i$, and moreover,
$N_I(v):=\bigcup_{i\in I}N_i(v)$ and $N_I(X):=\bigcap_{v\in X}N_I(v)$.
Provided that $G$ is equipped with linear order(s) $\leq^G$
on each $V_i$,~ $N_i^{\uparrow}(v)$ denotes the set of all
$w\in V_i$ such that there is $w'\in N_i(v)$ and $w'\leq^G\!w$
(all the vertices which are ``above'' some neighbour of $v$ in $V_i$),
and this is analogously extended to $N_I^{\uparrow}(v)$
and $N_I^{\uparrow}(X)$.

\begin{algo}
\label{alg:minmaxcliq}
\textsc{Input}:
An interval-monotone $k$-coloured clique instance $G$,
the colours classes $V(G)=V_1\,\dot\cup\dots\dot\cup\,V_k$
and the order $\leq^G$ on them.

\smallskip\textsc{Output}:
\textsc{Yes} if $G$ contains a clique of size $k$, and \textsc{No} otherwise.

\smallskip\textsc{Algorithm}:
Dynamically compute, for $i=2,3,\dots,k$,
sets $MinK^i(v)$ and $MaxK^i(v)$ where $v\in V_{i}$; such that
$MinK^i(v)$ is the $\leq^G$-minimum of all the cliques of size $i$ in $G$
which are contained in $\{v\}\cup V_1\cup\dots\cup V_{i-1}$
(note, these cliques must contain~$v$), or $\emptyset$ if nonexistent,
and $MaxK^i(v)$ is described analogously.

The computation of  $MaxK^{i}, MinK^{i}$ using values $MaxK^{2},\ldots ,
MaxK^{i-1}$ and values $MinK^{2},\ldots, MinK^{i-1}$ is described in the pseudocode
below.  Note that we have to compute both $MinK^i$ and $MaxK^i$ because we
compute $MinK^i$ from previously computed $MaxK^j$, $j<i$, and vice versa.
\begin{enumerate}[label=\arabic*.]
\item For every $v\in V_i$, set $X:=\{v\}$ and repeat:
\smallskip
\begin{enumerate}[label=\roman*)]
\item
 For $j=i-1,\dots,1$, and as long as $X\not=\emptyset$, do the following:
 \\
 find the minimum (wrt. $\leq^G$) element $x\in N_j(X)$ such that
 $j=1$ or $\emptyset\not=MaxK^j(x)\subseteq
 	N_{[1,j-1]}^{\uparrow}(X)\cup\{x\}$.
 If $x$ does not exist then $X:=\emptyset$, and otherwise set
 $X:=X\cup\{x\}$. Continue with next $j$.
\item
 Set $MinK^i(v):=X$.
\end{enumerate}
\smallskip
\item Analogously finish computation of $MaxK^i(v)$
using previous $MinK^j(x)$.
\smallskip
\item Output \textsc{Yes} if there is $v\in V_k$ such that $MinK^k(v)\not=\emptyset$,
and \textsc{No} otherwise.
\end{enumerate}
\smallskip
\end{algo}

\begin{thm}
Algorithm~\ref{alg:minmaxcliq} correctly solves any instance $G$
of interval-mono\-tone $k$-coloured \textsc{Multicoloured Clique} problem,
in time $O(k\cdot|E(G)|)$.
\end{thm}

\proof
It is enough to prove that the value of each $MinK^i(v)$ and $MaxK^i(v)$ is
computed correctly in the algorithm.
Let $K_{i,v}$ be the minimum of all the cliques of size $i$ in $G$
which are contained in $\{v\}\cup V_1\cup\dots\cup V_{i-1}$
(well-defined by Corollary~\ref{cor:minmaxcliq})---%
the correct value for $MinK^i(v)$.
Assume that some $MinK^i(v)=K_{i,v}'$ value is computed wrong, 
i.e., $K_{i,v}\not=K_{i,v}'$, and that $i$ is minimal among such wrong values.
Clearly, $i>2$.

If $K_{i,v}'=\emptyset$ then $K_{i,v}\not=\emptyset=K_{i,v}'$.
Otherwise we observe that, by the choices $x\in N_j(X)$ in step 1.a),
$K_{i,v}'\not=\emptyset$ is a clique of size $i$ in $G$ contained in 
$\{v\}\cup V_1\cup\dots\cup V_{i-1}$.
Consequently, $K_{i,v}'\not=\emptyset$ implies $K_{i,v}\not=\emptyset$, too.

Let $K_{i,v}''=K_{i,v}'$ if $K_{i,v}'\not=\emptyset$,
and otherwise let $K_{i,v}''$ be the last nonempty value of $X$ in the course of
computation of $MinK^i(v)$ in step 1.a) of the algorithm.
Since the tests in step 1.a) of the algorithm always succeed for
$x$ being $K_{i,v}\cap V_j$ and $X=K_{i,v}\cap(V_{j+1}\cup\dots\cup V_i)$,
there exists $j<i$ (and we choose such $j$ maximum) such that
$\{x\}=K_{i,v}\cap V_j\not=K_{i,v}''\cap V_j=\{x'\}$.
By the same argument, actually, $x>^G\!x'$.

Now, following iteration $j$ of step 1.a) of the algorithm
(which has ``wrongly'' chosen $x'$ instead of $x$),
let $K_0=MaxK^j(x')\cup\big(K_{i,v}\cap(V_{j+1}\cup\dots\cup V_i)\big)$.
The minimum of $K_{i,v}$ and $K_0$ is also a clique of size $i$,
by the interval-monotone property and Corollary~\ref{cor:minmaxcliq},
contradicting minimality of $K_{i,v}$ at~$x$.

In any case, indeed $K_{i,v}=K_{i,v}'$.
\smallskip

It remains to analyse the running time.
We consider separately every iteration of step 1,
each $v\in V_i$, for $i=2,\dots,k$.
Thanks to the interval-monotone property of $G$, we can preprocess
the neighbours of $v$ into subintervals of the classes $V_1,\dots,V_{i-1}$
with respect to $\leq^G$.
This is done in time $O\big(|N_{[1,i-1]}(v)|\big)$.
After that, every iteration $j$ of step 1.a) takes time 
$O\big(|N_{j}(v)|\cdot k\big)$, 
and so whole step~1 takes time $O\big(k\cdot|N_{[1,i-1]}(v)|\big)$.
Summing this over $v$ and $i$ as in the algorithm we arrive right at the estimate
$O(k\cdot|E(G)|)$.
\qed

\begin{cor}\label{cor:existfo-better}
  \textsc{Embedding} can be solved in time
  $O\big(\width(\PPP)^{|Q|}\cdot|Q|^3\cdot|P|^2\big)$.
\end{cor}
\proof
The reduction from embedding to compatible embedding has been shown within
Theorem~\ref{the:embedding-fpt}.
By the reduction here, $|V(G)|=O(|Q|\cdot|P|)$, $|E(G)|=|V(G)|^2$, and $k=|Q|$.
The runtime bound thus follows as in Theorem~\ref{the:embedding-fpt}.
\qed

\section{Kernelization Lower Bound}
\label{sec:lowerkernel}

Having shown that the \textsc{Embedding} problem is fixed-parameter
tractable, it becomes natural to ask whether it also allows for a
polynomial kernel. In this section we will show that this
unfortunately is not the case, i.e., we show that \textsc{Embedding}
does not have a polynomial kernel unless $\NOPOLYKERNEL$. 
Consequently, this also excludes a polynomial kernel for the \textsc{Poset FO-Model
  Checking} problem, of which \textsc{Embedding} is a special case. (\textsc{Poset FO-Model
  Checking} is an extension of \textsc{Poset $\exists$-FO-Model
  Checking} to the full FO logic.)

We will show our kernelization lower bound for \textsc{Embedding}
using the OR-composition technique outlined by
Proposition~\ref{pro:or-comp-no-poly-kernel}. Unfortunately, due to
the generality of the \textsc{Embedding} problem it turns out to be
very tricky to give an OR-composition algorithm directly for the
\textsc{Embedding} problem.
To overcome this problem, we introduce a restricted version of
\textsc{Embedding}, which we call \textsc{Independent Embedding}, for
which an OR-composition algorithm is much easier to find and whose
unparameterized version is still \NP-complete, as we prove below.

Let $\III_k=(I_k, \leq^{\III_k})$ be the poset that has $k$ mutually incomparable chains consisting of three
elements each. Then the \textsc{Independent Embedding} problem is
defined as follows.

\defparproblem{{\sc Independent Embedding}}{A poset
  $\PPP=(P,\leq^P)$ and a natural number
  $k$.}{$\width(\PPP)$, $k$}{Is there an embedding from $\III_k$ to
  $\PPP$?}\smallskip

\noindent \NP\hy completeness of \textsc{Independent Embedding} follows
straightforwardly from \NP\hy completeness of the ordinary independent set
problem on graphs.
As to an OR-composition algorithm for \textsc{Independent Embedding}, 
the other ingredient in Proposition~\ref{pro:or-comp-no-poly-kernel}, 
we do roughly as follows:
we first align a given collection of instances to the same (maximum) value
of the parameter~$k$, and then we ``stack'' these instances on top of one
another (all elements of a lower instance are ``$\leq^P$'' than all those of
a higher instance), 
making a combined instance of \textsc{Independent Embedding}
which is an OR-composition of all the input instances and whose
width does not exceed the maximum of their widths.
The formal proofs follow.\newpage

\begin{lem}
\label{lem:IEMBnpc}
  \textsc{Independent Embedding} is \NP\hy complete.
\end{lem}
\proof
  Since
  \textsc{Independent Embedding} is easily seen to be contained in \NP{},
  it suffices to show that it is \NP\hy hard. To show \NP\hy
  hardness we reduce from the well-known \textsc{Independent Set}
  problem in graphs, which given a graph $G$ and a natural number $k$,
  asks whether there are at least $k$ pairwise non-adjacent vertices in
  $G$. 
  For a graph $G$, we define the poset of $G$, denoted $\PPP_G=(P_G,\leq^{\PPP_G})$,
  as the poset having one chain $C_v$ consisting of three elements for each
  vertex $v$ of $G$, and where the bottom of the chain corresponding
  to a vertex $v$ is covered by the top of the chain corresponding to
  a vertex $u$ if, and only if, $\{u,v\} \in E(G)$.

  More formally,
  $\PPP_G$ has the elements $\SB a_v,b_v,c_v \SM v \in V(G) \SE$ and the
  relation $\leq^{\PPP_G}$ is defined by $x \leq^{\PPP_G}\! y$ if and only if 
  $x=y$, or $x=a_v$ and $y \in \{b_v,c_v\}$, or $x=b_v$ and $y=c_v$, or $x=a_v$
  and $y=c_u$ for some $u,v \in V(G)$ with $\{u,v\} \in E(G)$.
  Note that $\PPP_G$ is a poset, because $\leq^{\PPP_G}$ is acyclic and
  contains only the pairs given explicitly in the construction
  (i.e., there are no further arcs implied by transitivity since every $a_v$
  is a minimal element and every $c_v$ a maximal element), and that the
  only chains of length three in $\PPP_G$ are of the form $(a_v,b_v,c_v)$
  where~$v\in V(G)$.

  Then, for an instance $(G,k)$ of the \textsc{Independent Set} problem
  we construct the instance $(\PPP_G,k)$ of the \textsc{Independent
    Embedding} problem. Clearly, $(\PPP_G,k)$ can be constructed from
  $(G,k)$ in polynomial time, and if $G$ has an independent set of size at
  least $k$ then there is an embedding from $\III_k$ to $\PPP_G$. 
  Conversely, if $\III_k$ has an embedding into $\PPP_G$ then every
  length-$3$ chain of $\III_k$ is mapped into a distinct triple of the form
  $(a_v,b_v,c_v)$, where $v\in X\subseteq V(G)$ and $X$ is an independent
  set of size $k$ in $G$ since the distinct chains of $\III_k$ have
  mutually incomparable elements.
  This shows that \textsc{Independent Embedding} is \NP\hy complete.
\qed

\begin{lem}
\label{lem:IEMBnokernel}
  \textsc{Independent Embedding} does not have a polynomial kernel unless\\ $\NOPOLYKERNEL$.
\end{lem}
\proof
  To use the criterion of Proposition~\ref{pro:or-comp-no-poly-kernel}, we
  have got Lemma~\ref{lem:IEMBnpc} and now we need to show
  that there is an OR-composition algorithm for \textsc{Independent Embedding}.

  Suppose we are given $t$ instances
  $(\PPP_1,k_1),\dots,(\PPP_t,k_t)$
  of \textsc{Independent Embedding}. We first show that, w.l.o.g., we
  can assume that $k_1=\dotsb=k_t$. To see this let $k=\max_{1 \leq i
    \leq t}k_i$ and let $i$ with $1 \leq i \leq t$ be such that
  $k_i<k$. The idea is to replace every instance $(\PPP_i,k_i)$ with 
  the instance $(\PPP_i',k)$, where $\PPP_i'$ is the disjoint union of
  $\PPP_i$ and $\III_{k-k_i}$. Clearly, $(\PPP_i',k)$ is equivalent to
  $(\PPP_i,k_i)$ and can be constructed in polynomial time from
  $(\PPP_i,k_i)$. Furthermore, note that because
  $\width(\PPP_i')=\width(\PPP_i)+k-k_i$
  it also follows that $\width(\PPP_i')$ is
  bounded by $\width(\PPP_i)+k$.

  Hence, in the following we can assume that we are given $t$
  instances of \textsc{Independent Embedding} of the form
  $(\PPP_1,k),\dots,(\PPP_t,k)$. We will now construct a new (combined) instance
  $(\PPP,k)$ of \textsc{Independent Embedding} as follows. 
  The poset $\PPP=(P,\leq^P)$ is obtained from the disjoint union
  of the posets $\PPP_1,\dotsc,\PPP_t$
  after adding, for every $i$ and $j$ with $1 \leq i < j \leq t$,
  all the pairs $(p,p')$ such that $p \in P_i$ and $p' \in P_j$ to the
  ordering relation $\leq^P$. It follows from the construction that
  the width of $\PPP$ is equal to the maximum width of any $\PPP_i$. 
  Hence, the combined parameter $k+\width(\PPP)$ is
  bounded by (actually equal to) the maximum of the combined
  parameters of the instances $(\PPP_i,k)$. Furthermore, $(\PPP,k)$ can easily be constructed in
  time polynomial in $\sum_{1 \leq i \leq t}|\PPP_i|+k$. It thus
  only remains to show that $(\PPP,k)$ is a \textsc{Yes}-instance if and
  only if there is an $i$ with $1 \leq i \leq t$ such that $(\PPP_i,k)$
  is a \textsc{Yes}-instance.

  So suppose that $(\PPP,k)$ is a \textsc{Yes}-instance an let $e$ be an
  embedding from $\III_k$ to $\PPP$ witnessing this. W.l.o.g. we
  can assume that $k>1$ (because if $k=1$ we can solve each instance
  $(\PPP_i,k)$ in polynomial time, e.g., by going over all possible embeddings,
  and return a constant size \textsc{Yes}-instance if one of them is a
  \textsc{Yes}-instance and otherwise return a constant size
  \textsc{No}-instance). We claim that there is an $i$ with $1 \leq i
  \leq t$ such that $\SB e(q) \SM q \in I_k  \SE \subseteq
  P_i$. Suppose not then because $k>1$ there are $q$ and $q'$ in
  $I_k$ with $q \incom^{\III_k} q'$ such that $e(q)
  \in P_i$ and $e(q') \in P_j$ for some $i$ and $j$ with $1 \leq i < j
  \leq t$. It follows that $e(q) \leq^{P} e(q')$, which contradicts
  our assumption that $e$ is an embedding from $\III_k$ to $\PPP$
  since $q \incom^{\III_k} q'$. Hence, there is an $i$ with $1 \leq i
  \leq t$ such that $\SB e(q) \SM q \in I_k  \SE \subseteq
  P_i$. Consequently, $e$ is also an embedding from $\III_k$ to
  $\PPP_i$, as required.

  For the reverse direction suppose there is an $i$ with $1 \leq i
  \leq t$ such that $(\PPP_i,k)$ is a \textsc{Yes}-instance an let $e$ be an
  embedding from $\III_k$ to $\PPP_i$ witnessing this. Then $e$ is
  also an embedding from $\III_k$ to $\PPP$, as required.
\qed

We are now ready to summarize the main result of this section:

\begin{thm}
\label{the:nopolykernel-emb}
  \textsc{Embedding}, \textsc{Poset $\exists$-FO-Model Checking}
  and \textsc{Poset FO-Model Checking}
  have  no polynomial kernel unless $\NOPOLYKERNEL$.
\end{thm}
\proof
  The result for \textsc{Embedding} easily follows from the fact that
  \textsc{Independent Embedding} is a special case of the \textsc{Embedding}
  problem (and in particular there is a trivial polynomial parameter
  reduction from \textsc{Independent Embedding} to \textsc{Embedding}) and
  from Proposition~\ref{pro:poly-par-reduction-bi}. The \textsc{Poset
    $\exists$-FO-Model Checking} result is then easily proved
  by~Propositions~\ref{pro:equiv-FO-Embedding},
  and it is a special case of \textsc{Poset FO-Model Checking}. 
\qed

\section{Conclusions}
Besides establishing tractability of existential FO model checking on posets
of bounded width, the authors of~\cite{BGS14} also considered several other
poset invariants, giving (in-)tractability results for existential FO model
checking for these variants. This makes, together with our simplification of
proof of their main result, the parameterized complexity of the existential
FO model checking on posets rather well understood.

The main direction for further research, suggested already in~\cite{BGS14},
is the parameterized complexity of model checking of full FO logic on
restricted classes of posets, especially on posets of bounded width. This
problem is challenging, because currently known techniques for establishing
tractability of FO model checking are based on locality of FO and cannot be
applied easily to posets---transitivity of $\le$ causes that, typically, the
whole poset is in a small neighbourhood of some element. On the other hand,
attempts to evaluate an FO formula on a Hasse diagram
(i.e., on the graph of the cover relation of a poset) 
fail precisely because of locality of FO.

\bibliographystyle{alpha}
\bibliography{efo-lmcs}

\end{document}